\documentclass[]{raa}           

\usepackage{graphicx,times}
\usepackage{natbib}
\usepackage{amssymb,amsmath}
\bibpunct{(}{)}{;}{a}{}{,}
\usepackage{multirow}
\usepackage[section]{placeins}
\usepackage{float}
\usepackage{threeparttable}
\usepackage{subcaption}

\usepackage[pagebackref=true]{hyperref}
\usepackage{setspace} 

\begin{document}

   \title{Detection of TiO and VO in the atmosphere of WASP-121b and Evidence for its temporal variation}

 \volnopage{ {\bf 20XX} Vol.\ {\bf X} No. {\bf XX}, 000--000}
   \setcounter{page}{1}

   \author{Qinglin Ouyang
   \inst{1,3}, Wei Wang\inst{2}, Meng Zhai\inst{1}, Guo Chen\inst{4}, Patricio Rojo\inst{5}, Yujuan Liu\inst{2}, Fei Zhao\inst{2}, Jia-Sheng Huang\inst{1}, Gang Zhao\inst{2, 3}
   }

   \institute{ Chinese Academy of Sciences South America Center for Astronomy (CASSACA), National Astronomical Observatories, CAS, Datun Road A20, Beijing 100101, China; {\it mzhai@nao.cas.cn} \\
        \and
        CAS Key Laboratory of Optical Astronomy, National Astronomical Observatories, Chinese Academy of Sciences, Datun Road A20, Beijing 100101, China; {\it wangw@nao.cas.cn}\\
        \and
        School of Astronomy and Space Science, University of Chinese Academy of Sciences (UCAS), Yuquan Road A19, Beijing 100049, China \\
        \and 
        CAS Key Laboratory of Planetary Sciences, Purple Mountain Observatory, Chinese Academy of Sciences, Nanjing 210023, People’s Republic of China\\
        \and
        Departamento de Astronomía, Universidad de Chile, Camino El Observatorio 1515, Las Condes, Santiago, Chile
\vs \no
   {\small Received 20XX Month Day; accepted 20XX Month Day}
}

\abstract{We report the transit observations of the ultra hot Jupiter WASP-121b using the Goodman High Throughput Spectrograph (GHTS) at the 4-meter ground-based telescope Southern Astrophysical Research Telescope (SOAR), covering the wavelength range $502-900$\,nm. By dividing the target and reference star into 19 spectroscopic passbands and applying differential spectrophotometry, we derive spectroscopic transit light curves and fit them using Gaussian process framework to determine transit depths for every passbands. The obtained optical transmission spectrum shows a steep increased slope toward the blue wavelength, which seems to be too steep to be accounted for by the Rayleigh scattering alone. We note that the transmission spectrum from this work and other works differ obviously from each other, which was pointed out previously by \citet{Wilson2021} as evidence for temporal atmospheric variation. We perform a free chemistry retrieval analysis on the optical transmission spectra from this work and the literature HST/WFC3 NIR spectrum. We determine TiO, VO and H$_{2}$O with abundances of $-5.95_{-0.42}^{+0.47}$\,dex, $-6.72_{-1.79}^{+0.51}$\,dex, and $-4.13_{-0.46}^{+0.63}$\,dex, respectively. We compare the abundances of all these three molecules derived from this work and previous works, and find that they are not consistent with each other, indicating the chemical compositions of the terminator region may change over long timescales. Future multi-epoch and high-precision transit observations are required to further confirm this phenomena. We note that when combining the transmission spectra in the optical and in NIR in retrieval analysis, the abundances of V and VO, the NIR-to-optical offset and the cloud deck pressure may be coupled with each other. 
\keywords{methods: data analysis -- techniques: spectroscopic -- stars: individual (WASP-121) -- planetary systems -- planets and satellites: atmospheres.}
}

   \authorrunning{Qinglin Ouyang et al. }            
   \titlerunning{SOAR/GHTS transmission spectrum of WASP-121b}  
   \maketitle

%
\section{Introduction}           
\label{sec:intro}

Transmission spectroscopy is by far the most important technique in the study of exoplanet atmospheres. By determining the transit depths at different wavelength~\citep{Charbonneau2000, Deming2013, Madhusudhan_AR2019}, one can measure the absorption features of the planetary atmosphere and set constraints on the physical structure and chemical composition of the atmosphere at the day–night terminator region. Until now, many molecules and atoms, such as H$_2$O, CH$_4$, CO$_2$, TiO, VO, Na, K and Fe~\citep{Beaulieu2010, Giacobbe2021, Sedaghati2017, Casasayas-Barris2017, Chen2018, Ehrenreich2020}, have been discovered in the exoplanet atmospheres by transmission spectroscopy using large ground-based telescopes or space telescopes.

Ultra-hot Jupiters (UHJs) are a class of hot Jupiter-like planets with high temperatures ($\ge2000$\,K) and small distances to their host stars. They are excellent targets for transmission spectroscopy, and ideal laboratories for studying extreme chemistry and climates of exoplanets~\citep{Lothringer2018}. Such extremely high temperatures can make some refractory metals exist solely in their atomic states rather than molecule states~\citep{Hoeijmakers2018}, while the very close host star may photoionize some atoms in the upper atmosphere~\citep{Deibert2021}. On the other hand, UHJ with atmospheric circulation makes some atoms evaporate on the day side and condense on the night side, such as the ``iron rain'' in the atmosphere of WASP-76\,b, discovered by ~\citet{Ehrenreich2020}. Moreover, thermal inversion is frequently found on UHJs, due to the extra heating of the upper atmospheres by metal species  ~\citep{Lothringer2018, Gandhi2020}.
 
WASP-121b is an inflated UHJ discovered by ~\citet{Delrez2016} with $M_\textup{p} = 1.183~M_\textup{Jup}$ and $R_\textup{p}=1.865~R_{\textup{Jup}}$. Due to the high irradiation and short period ($T_\textup{eq} \ge 2400$\,K, $P = 1.275$\,days), it is one of the most well-studied UHJs using low-resolution (LR) transmission spectroscopy and high-resolution (HR) Doppler spectroscopy. ~\citet{Evans2016} first reported the H$_2$O detection and tentative TiO or VO absorptions from the near-infrared (NIR) transmission spectrum using the Hubble Space Telescope (HST). The following HST secondary eclipse observation confirmed the detection of H$_2$O, and revealed the thermal inversion in the atmosphere of WASP-121b, making it the first exoplanet found to contain a stratosphere~\citep{Evans2017}. However, none of the two subsequent secondary eclipse studies ~\citep{Mikal-Evans2019, Mikal-Evans2020} confirmed the existence of TiO and VO. The very recent spectroscopic phase curve observation of WASP-121b by \citet{Mikal-Evans2022} found that the temperature profile had changed from getting warmer with altitude on the dayside hemisphere to turning cooler with altitude on the nightside hemisphere. This effect may be significant enough to cause the condensation of metal oxides on the nightside, which may explain the none detection of TiO and VO. Using HR Doppler spectroscopy, an atomic library in the WASP-121b atmosphere was discovered by different instruments. ~\citet{Gibson2020} and \citet{Cabot2020} both detected Fe using UVES and HARPS spectrographs, respectively. Then the atoms Cr and V were found by ~\citet{Ben-Yami2020} using the archive HARPS data. These two atoms were then confirmed by ~\citet{Hoeijmakers2020}, which detected several new species including Mg, Na, Ca, and Ni. On the other hand, the hydrodynamic simulation of H$\alpha$ developed by ~\citet{Yan2021} suggested an expanding hydrogen envelope around this planet. The transmission spectrum obtained by Gemini/GMOS supported the possibility of temporal variation of the atmosphere of this particular planet~\citep{Wilson2021}. Recently, several HR abundance retrieval studies were performed successfully for this planet, including e.g., \citet{Gibson2022} and \citet{Maguire2023}, who reported consistent  relative abundance measurements of Fe, Cr and V based on the data from UVES and ESPRESSO, respectively.

Here we present the optical transmission spectrum of WASP-121b, obtained by the Goodman High Throughput Spectrograph \citep[GHTS, ][]{Goodman_intro2004} installed on the 4.1\,m Southern Astrophysical Research Telescope (SOAR), in Cerro Pachón, Chile. Recently, transmission spectrophotometry studies with 4-meter class telescopes becomes a new option in addition to the ground-based 6-10 meter telescopes or space telescopes, including the Low Resolution Ground-Based Exoplanet Survey using Transmission Spectroscopy (LRG-BEAST, \citealt{Kirk2016, Kirk2017, Kirk2018, Kirk2019, Alderson2020, Kirk2021}) and our recent work \citep{Ouyang2023}. 

In this work, we aim to investigate the presences of TiO/VO molecules or high-altitude clouds and hazes, which may provide us an opportunity to better understand the stratosphere of WASP-121b. This manuscript is organized as follows. We first introduce the observation and data reduction in Sec.~\ref{sec:obs and data}, and then present the light curve analysis in Sec.~\ref{sec:Light Curve Anlysis}. In Sec.~\ref{sec:Transmission Spectrum} we describe our retrieval analysis, followed by the discussion and conclusion in Sec.~\ref{sec:Discussion} and Sec.~\ref{sec:Conclusion}, respectively.

\section{Observation and Data reduction} \label{sec:obs and data}
\subsection{Observation} \label{subsec:Observation}

The observations of WASP-121b were taken on the nights of 2018 February 7 (hereafter Night 1) and 2018 March 1 (hereafter Night 2), using the Goodman High Throughput Spectrograph (GHTS) installed on SOAR (Program CN2018A-89, PI: Wei Wang). In order to monitor and correct the telluric variation, the Multi-Object Spectroscopy (MOS) mode of GHTS was used to simultaneously observe the target WASP-121 and the reference star HD\,55273 for differential spectrophotometry. The target and reference stars have similar $V$ magnitudes ~\citep[$10.51$ vs. $10.10$; ][]{Hog2000} and an angular separation of $\sim166.92\arcsec$.

The slit mask we used contains two pre-carved long wide slits, with one for the target star and the other for the reference star. The slits were set to have a projected width of 20$\arcsec$, to avoid slit loss that may occur in case of large seeing variation. The 400 M2 grating was used, which provides wavelength coverage 500-905\,nm and spectral resolution $R\sim$1850 with a seeing size of $\sim0.45\arcsec$. We employed the GG455 filter to block the second-order contamination and the $2\times2$ binning mode (0.30$\arcsec$ per binned pixel) for detector readout.

The two transit observations cover the UT window of $01:57-03:20$ of Night 1 and $00:53-06:06$ of Night 2, respectively. However, due to the dewpoint alert, at UT 03:21, the observation of Night 1 stopped about 1hr before the predicted mid-transit time T$_{\textup{mid}}$ = 04:39. During  Night 2, there is a small gap at starting from roughly $80$\,min after the transit center. This is because the target and reference star were slightly offset from the slit center and it took a few minutes to redo the acquisition. In total, we obtained 69 and 213 exposures for each night, with exposure times of $40-100$ seconds. The details of both observations are shown in Table~\ref{tab:observation details}. As only the Night 2 transit was covered entirely, the following data reduction procedure was only applied to the Night 2 data.

\begin{table*}
	\centering
	\caption{Details of the two transit observations.}
	\label{tab:observation details}
	\begin{threeparttable}
	\resizebox{\textwidth}{!}{
	\begin{tabular}{ccccccc} 
		\hline
		Observation night & Instrument & Start time (UTC) & End time (UTC) & Exposure time (s) & Exposure number & Airmass range\\
		\hline
		2018-02-07\tnote{a} & SOAR/GHTS & 01:57:22 & 03:21:34 & 40$-$60 & 69 & 1.03$-$1.01$-$1.02\\
		2018-03-01 & SOAR/GHTS & 00:53:59 & 06:06:52 & 60$-$100 & 213 & 1.02$-$1.01$-$1.95\\
		\hline
	\end{tabular}
	}
	\begin{tablenotes}
        \item[a] The data taken on 2018-02-07 is not used for following data analysis.
    \end{tablenotes}
    \end{threeparttable}
\end{table*}

\subsection{Data reduction} \label{sec:Data reduction}

The raw data set of Night 2 was reduced using \texttt{IRAF}~\citep{IRAF1986, IRAF1993} and the Python scripts written by us. We first corrected the overscan, bias, and flat field for the 2D spectra, and used the \texttt{dcr} package \citep{Pych2012} to remove cosmic ray hits. The 1D spectra extraction procedure was done by the optimal extraction algorithm developed by \citet{Horne1986}. To minimize the point-to-point scatter in the light curves, we also test a series of aperture widths from 17 pixels to 35 pixels. The aperture width of 32 pixels was finally chosen, which yielded the smallest out-of-transit scatter. For the wavelength calibration, the initial wavelength solution was derived using \texttt{identify} in \texttt{IRAF} from the HgArNe arc lamp spectra taken before and after science frames. Then the wavelength solution was refined by aligning all 1D spectra in the wavelength domain, using the strong telluric lines such as telluric O$_2$. The timestamp of each spectrum was set as the mid-exposure time, which is converted into Barycentric Julian Dates in Barycentric Dynamical Time (BJD$_{\textup{TDB}}$; \citealt{Eastman2010}).

\begin{figure}
	\includegraphics[width=\columnwidth]{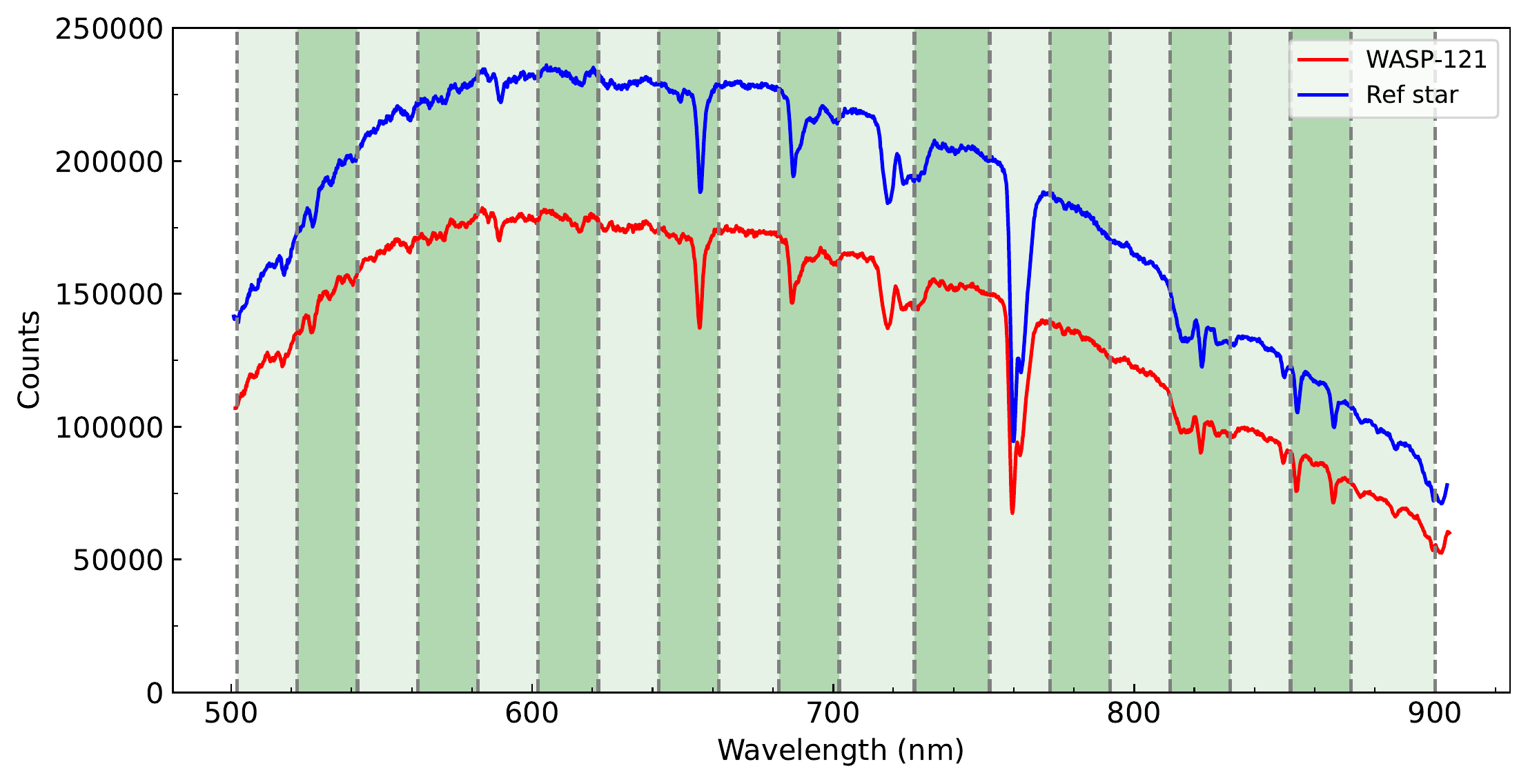}
    \caption{Example spectra of WASP-121 (red) and reference star (blue) for Night 2. The dark and light green shaded zones indicate the individual passbands used in this work.}
    \label{fig:star_spec}
\end{figure}

\begin{figure}
	\includegraphics[width=\columnwidth]{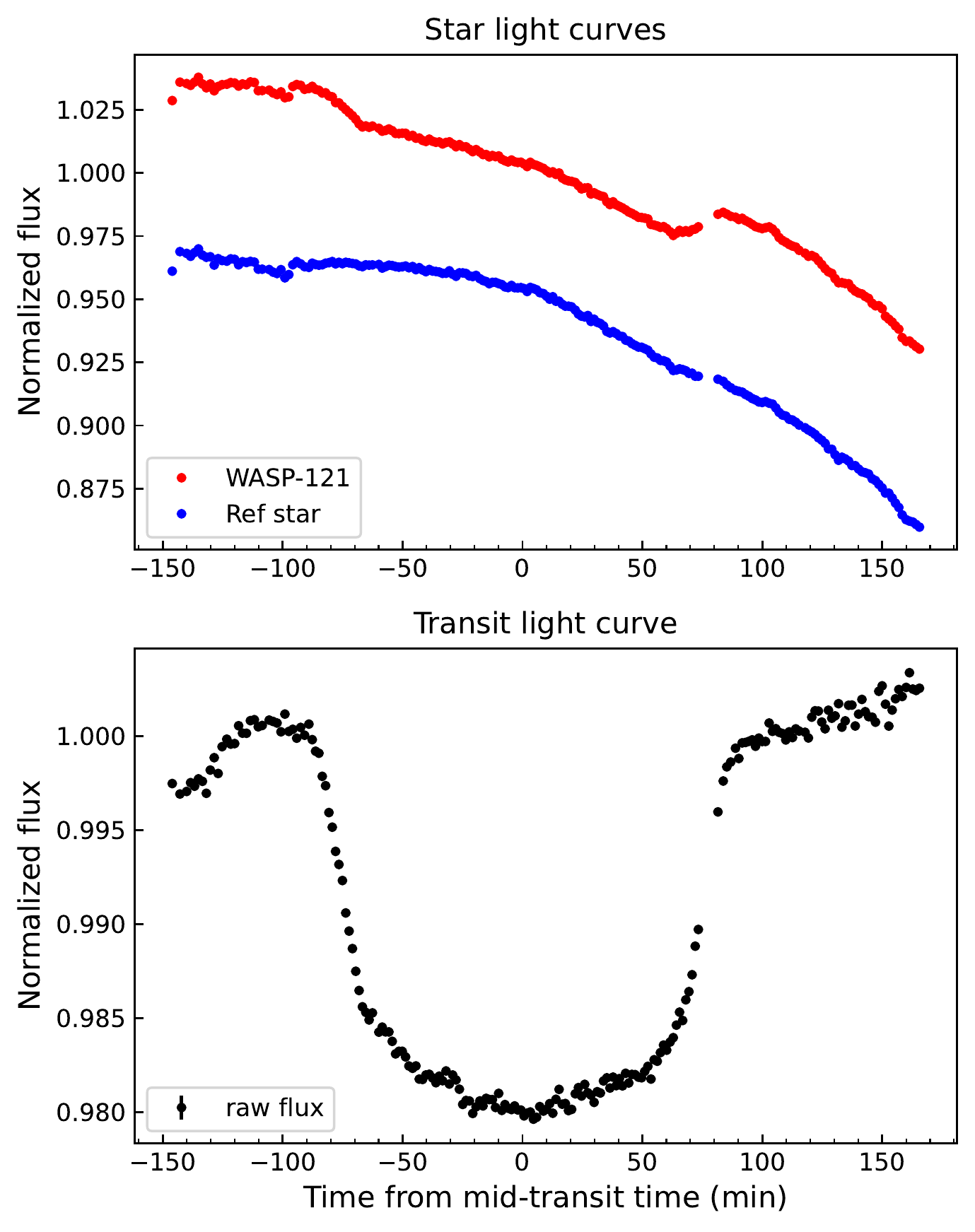}
    \caption{The normalized white light curves of Night 2. \textit{Top}: raw stellar light curves of WASP-121 (red) and reference star (blue) with the latter vertically shifted for clarity. Both light curves are normalized by the median value of out-of-transit flux}. \textit{Bottom}: the target-to-reference differential light curve normalized by the median value of out-of-transit flux.
    \label{fig:star_lightcurve}
\end{figure}

The white transit light curve was derived by differential spectrophotometry. We first integrated the fluxes within the wavelength range from 502\,nm to 900\,nm for both the target and reference star, and then divided the target star light curve by the reference star light curve, in order to remove the telluric and instrument effects. The spectroscopic transit light curves were created by a similar process for each of the pre-divided spectral channels. Following our experience in \citep{Ouyang2023}, we used 19 channels with 16 channels having widths of 20\,nm, 2 channels of 25\,nm (702$-$727\,nm and 727$-$752\,nm) and 1 of 28\,nm (872$-$900\,nm). Such division is to prevent the edge of each channel from falling on the prominent stellar absorption lines. Fig.~\ref{fig:star_spec} shows example spectra of the target and reference star of Night 2, along with the 19 spectral channels shown in the shaded light and dark green. The raw light curves of Night 2 for the target and reference star are shown in Figure~\ref{fig:star_lightcurve} in red and blue, respectively.

\section{Light Curve Analysis} \label{sec:Light Curve Anlysis}

\subsection{White light curve analysis} \label{sec:White light curve analysis}
The white light curve of Night 2 was generated by summing the flux between 502\,nm and 900\,nm. The derived light curves are considered to include two independent components, one component is the astrophysical transit signals and the other is noise. The former could be modeled by an analytic transit model ~\citep[e.g. the MA02 Model; ][]{Mandel2002} and noise, including the Gaussian noise (e.g. shot noise) and correlated systematics induced by telluric or instrument effects during the observation. Therefore, we treated the light curves as Gaussian process ~\citep[GP]{Rasmussen2006} with the MA02 model to find the systematics and transit depths. The GP method was firstly adopted in transit light curve analysis in ~\citet{Gibson2012(1)}, and then widely used in the study of transmission spectroscopy of exoplanets \citep{Gibson2013, Evans2015, Evans2017, Chen2021(1)} and our recent work by \citep{Ouyang2023}. It has the advantage of finding internal correlations between various parameters without providing specific functions and assessing how reliable these correlations are. In this section, we follow to use the GP method to analyze our light curves. The details are described in the following.

We treated the white light curve as a GP:
\begin{equation}
    f(\bm{t}, \bm{x}) \sim \mathcal{GP}(T(\bm{t},\bm{\phi}), \bm{\Sigma}(\bm{x}, \bm{\theta}))
	\label{eq:GP}
\end{equation}
where $T(\bm{t},\bm{\phi})$ is the transit function described by time series $\bm{\textit{t}}$ and transit parameters $\bm{\phi}$. $\bm{\Sigma}$ is the covariance matrix, which is a function of additional parameters $\bm{x}$ and hyperparameters $\bm{\theta}$ of GP. Additional parameters such as airmass, seeing, spectral trace movement in both the spectral and spatial directions vary with time during the observation, which may induce systematics. Correlations between data points are described using the covariance matrix, the element of which is given by:
\begin{equation}
    \bm{\Sigma}(x_\textnormal{n}, x_\textnormal{m}) = k(x_\textnormal{n}, x_\textnormal{m})
\end{equation}
where $k$ is the kernel function (or covariance function), which decides each element in the covariance matrix from additional parameters. In our case, we chose the Mat\'{e}rn $\nu$ = 3/2\, kernel which is widely used in transit light curve analysis ~\citep{Gibson2013, Chen2021(2), Chen2021(1)}. The kernel is defined as:
\begin{equation}
    k(x_\textnormal{n}, x_\textnormal{m}) = A(1+\sqrt{3}R_\textnormal{nm})\,\textnormal{e}^{-\sqrt{3}R_\textnormal{nm}}
\end{equation}
where $A$ and $R$ are the hyperparameters specifying the amplitude and the scale of the kernel, respectively.

We implemented GP via the Python package \texttt{george} \citep{Ambikasaran2015}, where the mean function of GP is the MA02 model implemented via the Python package \texttt{batman} \citep{Kreidberg2015}. We assumed a circular orbit with a fixed period derived by \citet{Delrez2016}. As for the additional parameters, we explored all the combinations of the parameters, e.g. time, airmass, seeing, and the position of the target star in the spatial ($x$) and dispersion direction ($y$), and then we chose the best parameter combination with the smallest Bayesian Information Criterion (BIC, ~\citealt{Schwarz1978}). As shown in Table~\ref{tab:BIC}, it turns out that employing time alone as the correlating parameter returns the minimal BIC and is therefore adopted in this work. 

\begin{table}
	\centering
	\caption{The BIC values calculated using the GP likelihood of different parameter combination as GP inputs. The best parameter combination is shown in bold.}
	\label{tab:BIC}
	\begin{threeparttable}
	\begin{tabular}{ccc} 
		\hline
		GP inputs & BIC & $\Delta$BIC \\
		\hline
		time, airmass, seeing, $x$, $y$ & -2798.96 & 22.08 \\
		time, seeing, $x$, $y$ & -2806.71 & 14.33 \\
		time, seeing, $x$ & -2809.40 & 11.64 \\
        time, seeing, $y$ & -2812.76 & 8.28\\
		time, $x$, $y$ & -2811.77 & 9.27 \\
		time, seeing & -2815.48 & 5.56 \\
		$\textbf{time}$ & $\textbf{-2821.04}$ & $\textbf{0}$ \\
		\hline
	\end{tabular}
    \end{threeparttable}
\end{table}

To account for the stellar limb-darkening (LD) effect on the transit light curve, the quadratic limb darkening law was adopted and the coefficients $u_{1}$ and $u_{2}$ were fitted using Gaussian priors. The priors were calculated by the Python package \texttt{PyLDTk} ~\citep{Parviainen2015}, which uses the library of PHOENIX-generated specific intensity spectra by \citet{Husser2013}. The stellar parameters reported by the discovery paper \citet{Delrez2016} were used, with the stellar effective temperature $T_\textnormal{eff}=6460\pm140$\,K, surface gravity log\,$g_{\star}=4.2\pm0.2$, and metallicity $Z=0.13\pm0.09$.

\begin{figure}
	\includegraphics[width=\columnwidth]{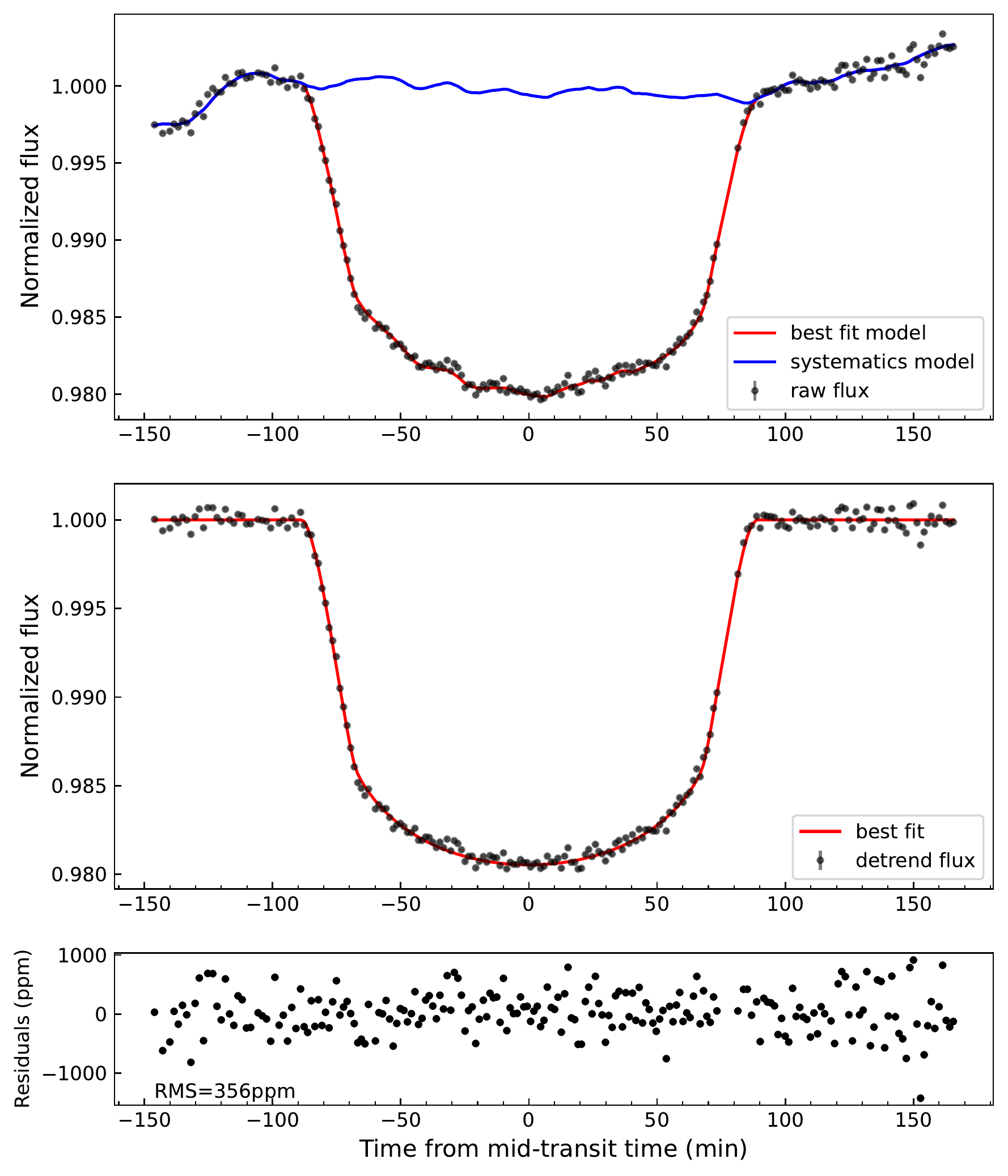}
    \caption{The raw, detrended light curves and residuals for Night 2 are shown from top to bottom successively. The black dots, red and blue lines represent the data, the best fit models, and the systematic model, respectively. The bottom panel shows the residuals obtained by subtracting the best fit model from the data, with an RMS of 356\,ppm.}
    \label{fig:white detrend lc}
\end{figure}

Then the Affine Invariant Markov Chain Monte Carlo (MCMC) was used via the Python package ~\texttt{emcee} ~\citep{Foreman-Mackey2013} to explore the marginalized posterior distributions of all fitting parameters. They are the mid-transit time $T_\textnormal{mid}$, the planet-to-star radius ratio $R_\textnormal{p}/R_{\star}$, the scaled semi-major axis $a/R_{\star}$, the orbit inclination $i$, limb-darkening coefficients $u_{1}$ and $u_{2}$, the hyperparameters of kernel function ($A$ and $R$) and the white noise jitter term $\sigma_\textnormal{w}$ which is to account for additional light-curve uncertainties. The period $P$ and the eccentricity $e$ were fixed to the values listed in \citet{Delrez2016}. Our MCMC procedure consisted of 100 walkers, each one with 20000 steps, and the first 4000 steps were set as the "burn-in" phase. The chain length is kept to be more than 50 times of autocorrelation time to ensure convergence. The white light curve and the best-fit model are shown in Fig.~\ref{fig:white detrend lc}. The priors and posterior distributions of the fitting parameters are listed in Table~\ref{tab:fitted parameters}.

\begin{table}
\renewcommand\arraystretch{1.5}
	\centering
	\caption{The priors and posterior results for the white light curve analysis.}
	\label{tab:fitted parameters}
	\begin{threeparttable}
	\begin{tabular}{lcc} 
		\hline
		\hline
		Parameters & Prior & Posterior\\
		\hline
		Period [days] & $1.2749255^{a}$  & ---\\
		\textit{e} & $0^{a}$ & ---\\
		\hline
		$T_\textnormal{mid}$ [$\textnormal{MJD}^{b}$] & $\mathcal{U}(8179.60, 8179.70)$ & $8179.64181_{-0.00027}^{+0.00028}$\\
		$R_\textnormal{p}/R_{\star}$ & $\mathcal{U}(0.10, 0.20)$ & $0.12772_{-0.00322}^{+0.00229}$\\
		$a/R_{\star}$ & $\mathcal{U}(0, 5)$ & $3.74_{-0.03}^{+0.03}$\\
		$i$[deg] & $\mathcal{N}(87.6, 0.6)$ & $87.28_{-0.58}^{+0.60}$\\
		$u_{1}$ & $\mathcal{N}(0.439, 0.006)$ & $0.43950_{-0.00543}^{+0.00544}$\\
		$u_{2}$ & $\mathcal{N}(0.139, 0.008)$ & $0.13917_{-0.00864}^{+0.00853}$\\
		ln\,$A$ & $\mathcal{U}(-20, -1)$ & $-12.96_{-0.60}^{+0.84}$ \\
        ln\,$R_{t}$ & $\mathcal{U}(-15, 15)$ & $-7.23_{-0.68}^{+0.82}$ \\
		ln\,$\sigma_\textnormal{w}$ & $\mathcal{U}(-20, -1)$ & $-15.89_{-0.12}^{+0.12}$\\
		\hline
	\end{tabular}
	\begin{tablenotes}
        \item[a] These parameters were fixed in light curve modelling, using the values from \citet{Delrez2016}.
        \item[b] MJD = BJD$_{\textup{TDB}}$ - 2450000.
    \end{tablenotes}
    \end{threeparttable}
\end{table}

\subsection{Spectroscopic light curves analysis} \label{sec:Spectroscopic light curve analysis}

\begin{figure*}
	\includegraphics[width=\textwidth]{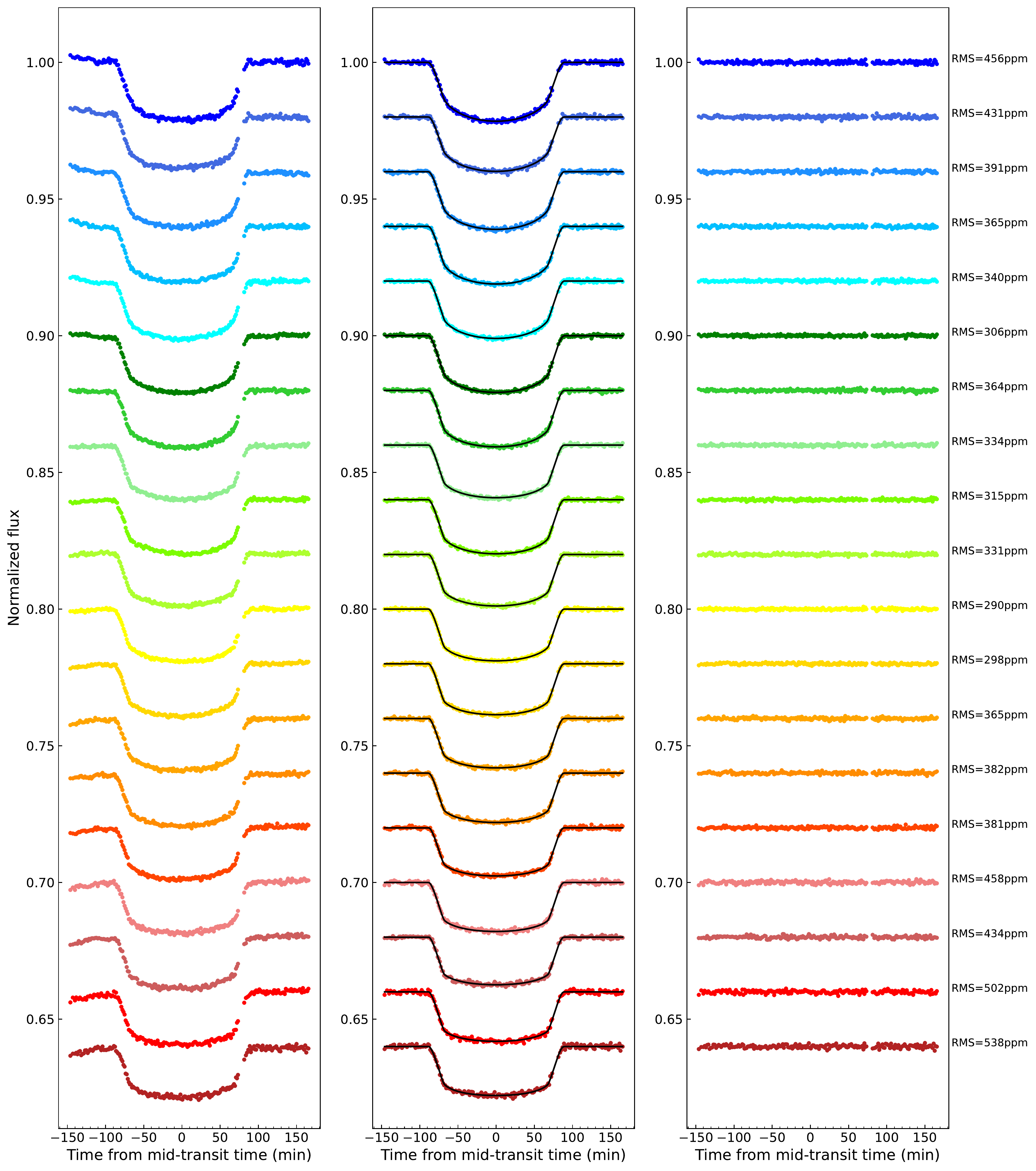}
    \caption{The 19 spectroscopic light curves obtained for Night 2. For distinction, each light curve was applied a constant vertical offset and a different color. \textit{Left panel}: the raw light curves with common-mode noise removed. \textit{Middle panel}: the detrended light curves (colored dot) with the best-fit model curves (black lines). \textit{Right panel}: the observation$-$model residuals.}
    \label{fig:passband lc}
\end{figure*}

The ``raw'' spectroscopic light curves were created by summing the flux in every channel described in Sec.~\ref{sec:obs and data}. They were divided by common-mode noise which had been derived previously by dividing the white light curve to its best-fit transit model, i.e., the ``divide-white'' method which is widely used and is proven to be able to provide significant improvements to the precision of the transit depth measurements. Next, the light curve in each channel was modelled separately to derive the wavelength-dependent parameters, which are $R_\textnormal{p}/R_{\star}$, $u_{1}$ and $u_{2}$. The wavelength-independent parameters ($T_\textnormal{mid}$, $a/R_{\star}$, $i$) with the best-fit values are derived from the white light curve fitting (Table.~\ref{tab:fitted parameters}). The number of MCMC walkers was set to be 50 and each walker contain 20000 steps, with the first 4000 steps being the ``burn-in'' phase. All the 19 spectroscopic light curves with their best-fit models are shown in Figure~\ref{fig:passband lc}.

\section{Transmission Spectrum} \label{sec:Transmission Spectrum}

\begin{table}
\renewcommand\arraystretch{1.5}
    \centering
    \caption{The planet-to-star radius ratios of all passbands of Night 2.}
    \label{tab:Rp/Rs}
    \begin{threeparttable}
    \begin{tabular}{ccc} 
        \hline
        Wavelength Center (nm) & Width (nm) & $R_\textup{p}/R_{\star}$ \\
        \hline
        512.0 & 20 & $0.13058_{-0.00097}^{+0.00100}$ \\
        532.0 & 20 & $0.12601_{-0.00293}^{+0.00375}$ \\
        552.0 & 20 & $0.13031_{-0.00203}^{+0.00228}$ \\
        572.0 & 20 & $0.13079_{-0.00211}^{+0.00236}$ \\
        592.0 & 20 & $0.13096_{-0.00189}^{+0.00140}$ \\
        612.0 & 20 & $0.13060_{-0.00120}^{+0.00097}$ \\
        632.0 & 20 & $0.13079_{-0.00089}^{+0.00065}$ \\
        652.0 & 20 & $0.12738_{-0.00059}^{+0.00058}$ \\
        672.0 & 20 & $0.12879_{-0.00089}^{+0.00093}$ \\
        692.0 & 20 & $0.12599_{-0.00075}^{+0.00113}$ \\
        714.5 & 25 & $0.12665_{-0.00113}^{+0.00076}$ \\
        739.5 & 25 & $0.12604_{-0.00121}^{+0.00119}$ \\
        762.0 & 20 & $0.12445_{-0.00170}^{+0.00168}$ \\
        782.0 & 20 & $0.12467_{-0.00145}^{+0.00143}$ \\
        802.0 & 20 & $0.12328_{-0.00183}^{+0.00173}$ \\
        822.0 & 20 & $0.12481_{-0.00192}^{+0.00174}$ \\
        842.0 & 20 & $0.12346_{-0.00287}^{+0.00267}$ \\
        862.0 & 20 & $0.12610_{-0.00234}^{+0.00222}$ \\
        886.0 & 28 & $0.12496_{-0.00296}^{+0.00274}$ \\
        \hline
    \end{tabular}
    \end{threeparttable}
\end{table}

\begin{figure}
	\includegraphics[width=\columnwidth]{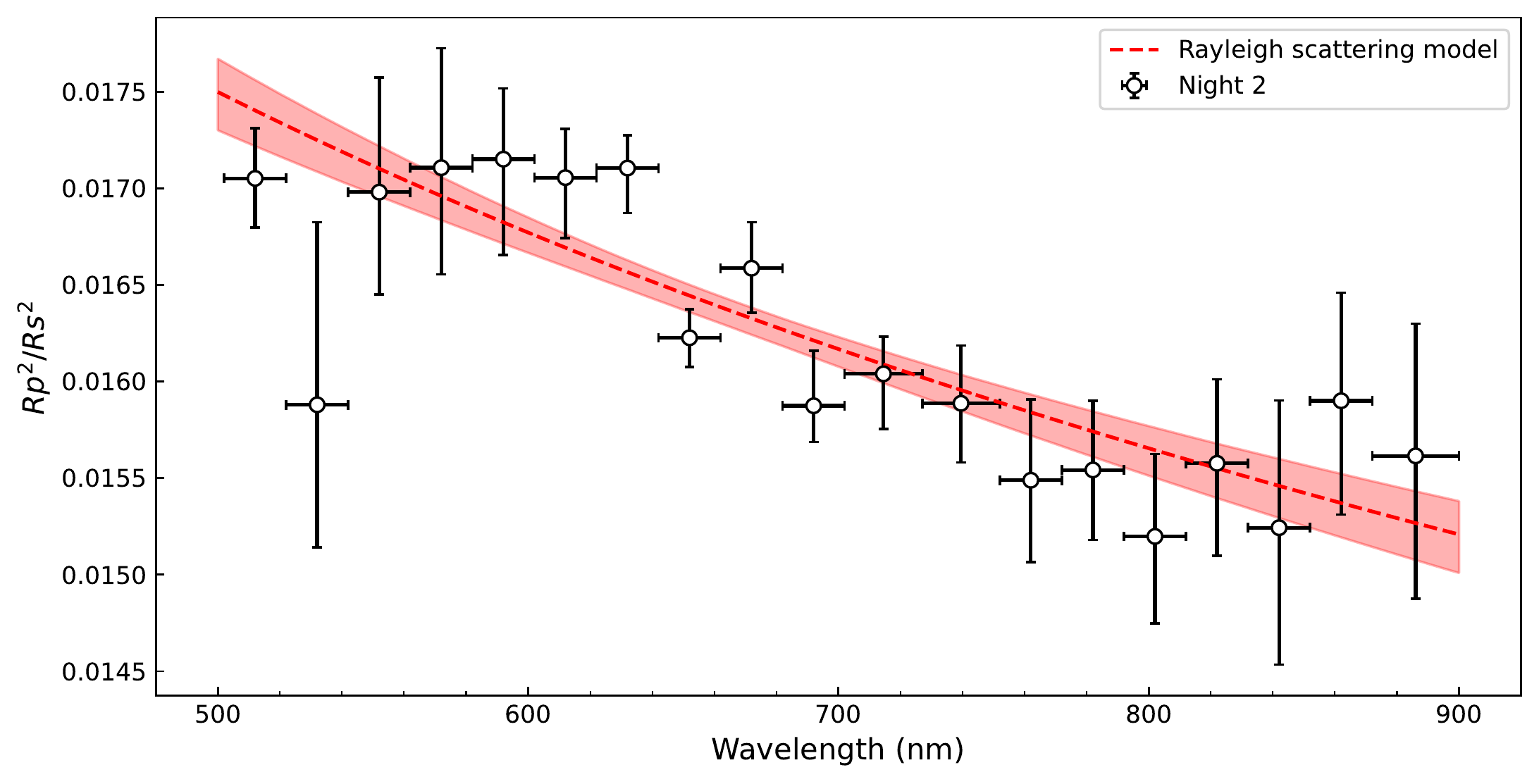}
    \caption{The SOAR/GHTS optical transmission spectrum obtained in this work with a Rayleigh scatting model, red shaded area is the 1$\sigma$ confidence interval of the model.}
    \label{fig:SOAR_tran_spec}
\end{figure}

\subsection{The transmission spectra} \label{subsec:discrepancy}

The derived $R_\textnormal{p}/R_{\star}$ of all the 19 passbands are listed in Table~\ref{tab:Rp/Rs}, and the corresponding transmission spectrum is shown in Fig.~\ref{fig:SOAR_tran_spec}. As described in Section\ref{sec:intro}, the transmission spectra of this planet have been obtained by several previous works, using different ground-based and space instruments~\citep[][or D16, E16, E18, W21 in short, respectively]{Delrez2016, Evans2016, Evans2018, Wilson2021}. It is well worth comparing our obtained SOAR optical spectrum (hereafter the OPT data) with them for consistency check and to explore the evidence of temporal changes of exoplanetary atmospheres. 

\begin{figure*}
	\includegraphics[width=\textwidth]{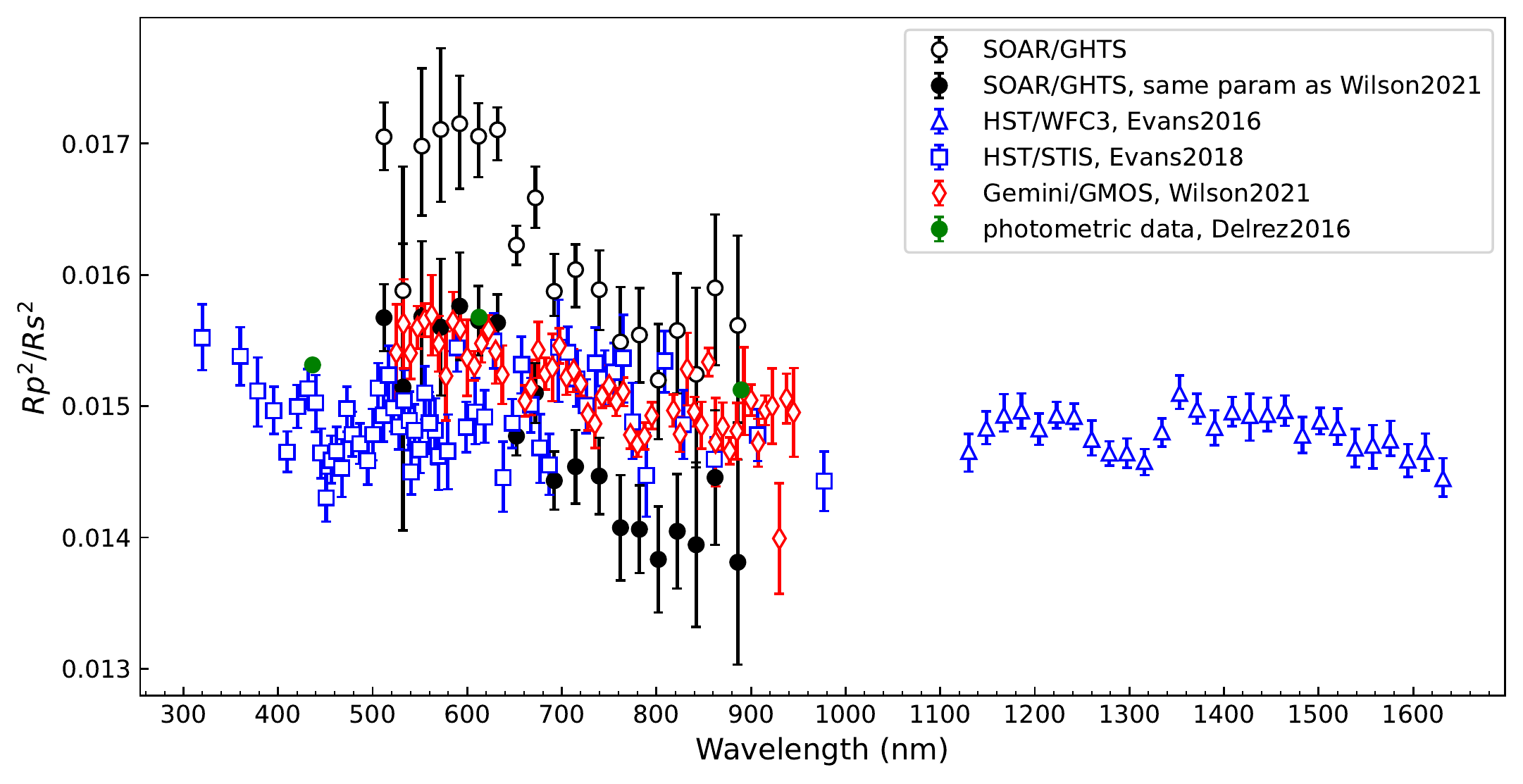}
    \caption{The optical and NIR transmission spectra obtained in this work and previous works.}
    \label{fig:tran_spec_comp}
\end{figure*}

We show in Figure~\ref{fig:tran_spec_comp} the transmission spectra from the OPT data and from D16, E16, E18 and W21. It is obvious from the plot that OPT (the black open circles) shows a large discrepancy from the others, with a relatively larger effective planet radius and steeper slope toward blue.
The discrepancy may be caused by slightly different input parameters such as $a/R_{\star}$ and $i$. Note that W21 employed the orbital parameters the same as the reported value in E18 (i.e. $a/R_{\star} = 3.86$, $b = 0.06$) and using a Gaussian prior centered on the E18 value for $R_\textup{p}/R_{\star}$, i.e., $\mathcal{N}(0.1219, 0.0005)$, which are different from ours as listed in Table~\ref{tab:fitted parameters}. It is pointed out by \citet{Alexoudi2018} and \citet{Alexoudi2020} that for employing different input orbital parameters may induce  notable discrepancy of transmission spectra, which is likely to be a steeper slop for the cases of K- and F- type stars or a feature at blue wavelength for M-type stars. The difference between the SOAR and W21 Gemini transmission spectra, i.e., $\Delta R_\textnormal{p}^{2}/R_{\star}^{2}$, is $\sim 1453$\,ppm. However, the discrepancy still remains albeit a bit smaller, with a $\Delta R_\textnormal{p}^{2}/R_{\star}^{2} \sim 1216$\,ppm, when we use the orbital parameters exactly the same as in W21 to redo the SOAR white light curve fitting. Only if a stringent Gaussian prior like the W21 one is adopted, the offset can turn close to zero. Even in such a case, the shape of the newly derived SOAR transmission spectrum remains obviously different from the W21 spectrum, as shown by the filled circles in Fig.~\ref{fig:tran_spec_comp}. We thus conclude that using different orbital parameters should not be the major cause for the observed discrepancy between the three optical transmission spectra in WASP-121b's case.

There is a possibility that the data analysis processes used in this work lead to biased results. To examine this possibility, we performed a new data analysis on the two-night Gemini/GMOS 1D spectra provided by the authors of W21 via private communication and derived a ``new'' GMOS transmission spectra. The white and spectroscopic light curves were created mainly following the procedures described in Section~2 of W21 using our pipeline. For the light curve analysis, we followed the same procedure described in Sec.~\ref{sec:Light Curve Anlysis}. We treated the priors of all fitting parameters the same as did in W21 and fitted two-night data jointly using our Python scripts. The new transmission spectrum we obtained from the GMOS data is consistent with the published one in W21 (Fig.~\ref{fig:tran_spec_gmos}), with a small offset $\sim472 \pm 286 \pm 25 $\,ppm. 

The small offset between the two spectra is not surprising because our data analysis process is not exactly the same as those in W21. For example, in the white light curve analysis, they fixed orbital parameters to the value from E18, i.e., $a/R_{\star} = 3.86$, $b = 0.06$, and set an extremely small Gaussian prior to $R_\textup{p}/R_{\star}$, i.e. $\mathcal{N}(0.1219, 0.0005)$. In our case, we fix $a/R_{\star} = 3.86$, the inclination $i = 89.1^{\circ}$ and $R_\textup{p}/R_{\star}\sim \mathcal{U}(0.1, 0.2)$. The value of $i$ is set to correspond to $b=0.06$ at the given $a/R_{\star}$. So, the only difference in parameter setting is the prior of $R_\textup{p}/R_{\star}$ - W21 used a narrow Gaussian distribution so that $R_\textup{p}/R_{\star}$ could hardly be changed, while we adopted a much looser uniform distribution.

More importantly, these two spectra share the same overall trend and distribution, as indicated by the linear fittings and the K-S test. The linear fittings are performed between the obtained transit depths and the wavelength in nanometers of the two transmission spectra, which yield quite similar linear coefficients ($-1.92 \times 10^{-6}$ of ``new'' GMOS vs. $-1.99 \times 10^{-6}$ of W21), and the K-S test implies that these two data points probably originate from the same probability distribution at a confidence level of 95\% with a quite small D statistic of $0.094$. Therefore, we conclude that the data analysis procedures employed in this work do not lead to any significant bias or trend, and the observed discrepancies between this work and all the others are likely to be real, i.e., there may be temporal changes in the atmosphere. This is supported by the fact that all the other three transmission spectra are different from each other as well, having both different mean values and slopes.



\subsection{The steep slope in the transmission spectrum}
\label{subsec:slope}
The steep increased slope toward the blue wavelength is shown in Fig.~\ref{fig:SOAR_tran_spec}, which we first consider as the result of the Rayleigh scattering in the atmosphere of WASP-121b. \citet{Lecavelier2008} pointed out that the measured effective planet radius related to Rayleigh scattering should be a function of the wavelength, which is given by:
\begin{equation}
\centering
    \frac{\mathrm{d}R_p}{\mathrm{d}\ln{\lambda}} = \alpha H = \alpha\frac{k_B T}{\mu g}
	\label{eq:scattering slope equation}
\end{equation}
where $H$ is the scale height of the planet atmosphere, $T$ is the planetary temperature, $k_\textnormal{B}$ is the Boltzmann constant, $\mu$ is the mean mass of atmospheric particles, and $g$ is the gravity of the planet. According to the derivation in Sec.~3 of \citet{Lecavelier2008}, the value of $\alpha$ caused by Rayleigh scattering should be close to $-4$.

We fit the transmission spectrum using Equation~\ref{eq:scattering slope equation} to determine the value of $\alpha$, with $T_\textnormal{eq} = 2358\,K$, log\,$ g = 2.973$ (cgs) taken from Table~4 in~\citet{Delrez2016}, and the mean mass of atmospheric particles $\mu = 2.37$ assuming a hydrogen-dominated atmosphere. We derive $\alpha = -9.95_{-1.49}^{+1.47}$, which is much smaller than the predicted value $\alpha = -4$ if only the Rayleigh scattering effect is concerned. This suggests that the Rayleigh scattering should not be the only cause for the observed steep slope in the transmission spectrum.

\subsection{Retrieval analysis}

To further evaluate the atmospheric composition and temperature structure of WASP-121b, we perform retrieval analysis on the derived transmission spectra. We use the \texttt{petitRADTRANS} package~\citep{Molliere2019} and the \texttt{PyMultiNest} package~\citep{Buchner2014} to generate 1D model transmission spectrum and calculate the Bayesian evidence($\mathcal{Z}$). We adopt the two-point (2P) temperature-pressure ($T-P$) profile given by~\citet{Brogi2014} and assume free chemistry to find the specific species causing the features in the transmission spectrum. For the 2P $T-P$ profile, the temperatures are constant at altitudes above the lower pressure point ($T_1$, $P_1$) or below the higher pressure point ($T_2$, $P_2$), respectively, and change linearly with log\,$P$ between the two points. The parameters ($T_{1,2}$, $P_{1,2}$) are free parameters to be retrieved. Although a dozen of atoms and molecules has been discovered in the atmosphere of WASP-121b, it is not realistic to retrieve all of them with our LR data obtained on a ground-based 4m telescope. Therefore, we only take a few species including H$_2$O, VO, TiO, FeH, Na, Ca, Fe, Mg, and V, which have relatively strong absorption in the optical and near-infrared. Molecule H$_2$ and He are treated as filling gases with atomic abundance ratios fixed to the solar values, and their Rayleigh scattering are considered in the model spectra. The mass fractions of the other 9 species are allowed to vary freely. In addition, the reference planet radius $R_{p}$ is set as a free parameter, while the reference pressure $P_{0}$ is fixed to be 0.01\,bar. 

We perform in total three sets of retrieval analysis to explore the bulk composition of the atmosphere of WASP-121b. The first set is to find the most contributing species among H$_2$O, VO, TiO and FeH and determine their abundances, based on the combination of our OPT data and the literature near-infrared transmission spectrum from \citet{Evans2016} (hereafter the ONIR data). We choose these four molecules because H$_2$O, VO, \ion{Fe}{I} and \ion{Fe}{II} have been firmly detected, while TiO have been tentatively detected~\citep{Evans2016, Evans2018, Sing2019, Gibson2020}. We run 4 retrieval experiments to determine the best-fit of model with different combinations of the four molecules (cf. Models A$1-4$ in Table~\ref{tab:retrieval_statistics}). We find that Model 2 with H$_2$O, VO, TiO included matches best with the ONIR data, given its largest Bayesian evidence ln\,$\mathcal{Z}$ and the smallest reduced chi-square $\chi_{\nu}^{2}$. We note that this model is significantly better than the two models with TiO excluded, i.e., Models A1 and A3. It is also evident that adding FeH to the A2 model does not return better results, suggesting that the ONIR data does not support a significant presence of FeH in WASP-121b's atmosphere. 

Next, we run another 5 models with each one the combination of the A2 model and one of the following atoms, Mg, Na, Ca, Fe and V to explore the presence of them. The results are shown as Models B$1-5$ in Table~\ref{tab:retrieval_statistics}. We find that B5 gives the best fits to the data set with $\chi_{\nu}^{2}$ smaller than the A2 model, while the other models have smaller ln\,$\mathcal{Z}$ and larger $\chi_{\nu}^{2}$. This implies that WASP-121b's atmosphere may possess a large amount of atom V. Finally, based on B5 model, we perform retrieval analysis on the OPT data alone. This is to explore what we can infer if only data in the optical band is available. Unfortunately, the retrieval is not very robust given that the best-fit model returns a large $\chi_{\nu}^{2}$ of 2.65.

\begin{figure*}
	\includegraphics[width=\textwidth]{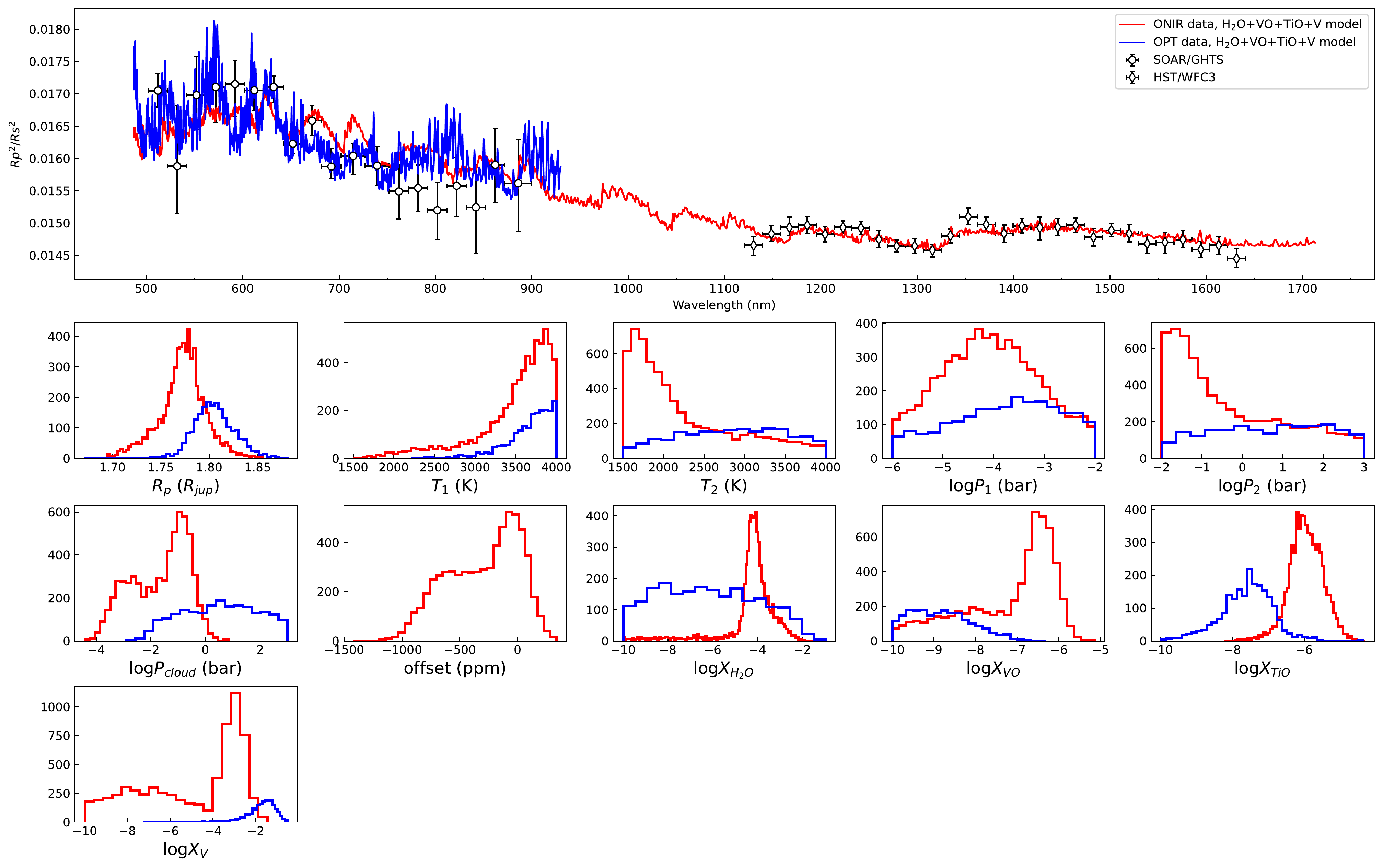}
    \caption{The OPT and ONIR transmission spectra of WASP-121b with best-fit models overplotted. \textit{Black open circles with error bars}: the SOAR optical transmission spectrum derived in this work. \textit{Black open diamonds with errorbars}: the HST/WFC3 NIR transmission spectrum from~\citet{Evans2016}. The blue and red solid lines are the best-fit retrieval models for OPT and ONIR data, respectively.}
    \label{fig:tran_spec_free_model}
\end{figure*}

The statistics of the retrievals mentioned above are summarized in Table~\ref{tab:retrieval_statistics}. The best-fit models of the ONIR (the B5 model) and the OPT data (the C1 model) are shown in Figure~\ref{fig:tran_spec_free_model}, and the corresponding posterior distributions are listed in Table~\ref{tab:retrieval}. It is noted that the derived planet radius from the OPT data is consistent with that from the ONIR data within 1$\sigma$, although a little bit larger. As illustrated in Figure~\ref{fig:TP_profile}, the presence of thermal inversion is favored by both retrievals, despite that the derived $T-P$ profiles are not well consistent with each other. Given that the optical transmission spectrum is from a ground-based telescope, while the NIR spectrum is from a space telescope, there might be an offset between these two datasets as noted previously by \citet{Murgas2020}. Therefore, we add $\delta$, the offset between the HST/WFC3 data and our SOAR data, as a free parameter for the ONIR retrievals.

The ONIR data yields a $\sim14.2\sigma, 6.6\sigma$ and $3.8\sigma$ detection of TiO, H$_2$O and VO, respectively. In addition, the atom V is detected with an abundance of $-3.77_{-4.21}^{+1.03}$, suggesting that atom V may present in WASP-121b's atmosphere. However, the ln\,$\mathcal{Z}$ value of the B5 model is only 0.9 larger than the A2 model, and there is a long tail extending to lower abundance in the posterior distribution plot in Figure~\ref{fig:tran_spec_free_model}. Therefore, we can only claim a tentative detection of V with the abundance upper limit as $< -2.74$ in this work.
For comparison, the retrieval analysis on the OPT data is not ideal, with only TiO and VO reasonably well constrained with log\,$X_\textnormal{TiO} = -7.54_{-0.74}^{+0.60}$ and log\,$X_\textnormal{VO} = -8.89_{-0.68}^{+0.82}$, respectively. However, the derived V abundance is too high to be real, which may be the cause for the relatively smaller abundances of TiO and VO with respect to those from the ONIR data. We suggest that the worse precision of the optical transmission spectrum hinders the accurate constraint of some species and $T-P$ profile. To conclude, the B5 model yields the best result. The retrieval models and the posterior distributions plots of the A and B series models are shown in Figure~\ref{fig:tran_spec_model_A} and Figure~~\ref{fig:tran_spec_model_B}, respectively.

\begin{table*}
	\centering
	\caption{Statistics of all retrieval models}
	\label{tab:retrieval_statistics}
	\begin{threeparttable}
	\begin{tabular}{lccccccccc} 
		\hline
	    $\#$ & Model & \multicolumn{4}{c}{OPT} & \multicolumn{4}{c}{ONIR} \\
	    \hline
	     & & dof & $\chi_{\nu}^{2}$ & ln\,$\mathcal{Z}$ & $\Delta$ ln\,$\mathcal{Z}$ & dof & $\chi_{\nu}^{2}$ & ln\,$\mathcal{Z}$ & $\Delta$ ln\,$\mathcal{Z}$ \\
	    \hline
        \multicolumn{10}{l}{\bf Only molecules} \\
	    \noalign{\smallskip}
	    A1 & H$_2$O $+$ VO &  &  &  &  & 38 & 2.12 & 397.7 & -16.5 \\
	    A2 & H$_2$O $+$ VO $+$ TiO &  &  &  &  & 37 & 2.02 & 414.2 & 0 \\
        A3 & H$_2$O $+$ VO $+$ FeH &  &  &  &  & 37 & 2.18 & 397.1 & -17.1 \\
        A4 & H$_2$O $+$ VO $+$ TiO $+$ FeH &  &  &  &  & 36 & 2.03 & 413.0 & -1.2 \\
		\hline
		\noalign{\smallskip}
		\multicolumn{10}{l}{\bf A2 + one atom} \\
		\noalign{\smallskip}
		B1 & A2 $+$ Ca &  &  &  &  & 35 & 2.02 & 413.6 & -1.5 \\
		B2 & A2 $+$ Fe &  &  &  &  & 35 & 2.07 & 414.1 & -1.0 \\
		B3 & A2 $+$ Mg &  &  &  &  & 35 & 2.06 & 414.0 & -1.1 \\
        B4 & A2 $+$ Na &  &  &  &  & 35 & 1.99 & 413.4 & -1.7 \\
        B5 & A2 $+$ V  &  &  &  &  & 35 & 1.55 & 415.1 & 0\\
		\hline
		\noalign{\smallskip}
		\multicolumn{10}{l}{\bf B5 on the OPT data} \\
		\noalign{\smallskip}
		C1 & B5 & 8 & 2.65 & 109.2 & 0 &  &  &  & \\
		\hline
	\end{tabular}
	\end{threeparttable}
\end{table*}

\begin{table*}
\renewcommand\arraystretch{1.5}
	\centering
	\caption{The posterior distribution of all retrieval parameters from best-fit models of OPT and ONIR data.}
	\label{tab:retrieval}
	\begin{threeparttable}
	\begin{tabular}{cccc} 
		\hline
	    Free chemistry & Prior & Posterior of OPT & Posterior of ONIR \\
		\hline
		$R_\textnormal{p}$ ($R_\textnormal{jup}$) & $\mathcal{U}(1.6, 2.0)$ & $1.803_{-0.019}^{+0.022}$ & $1.774_{-0.024}^{+0.020}$ \\
        $T_{1}$ (K) & $\mathcal{U}(1500, 4000)$ & $3726_{-311}^{+196}$ & $3626_{-545}^{+252}$ \\
		log\,$P_{1}$(bar) & $\mathcal{U}(-6, -2)$ & $-3.67_{-1.27}^{+1.02}$ & $-4.08_{-1.01}^{+1.03}$ \\
        $T_{2}$ (K) & $\mathcal{U}(1500, 4000)$ & $2844_{-750}^{+693}$ & $2004_{-351}^{+1030}$ \\
        log\,$P_{2}$(bar) & $\mathcal{U}(-2, 3)$ & $0.72_{-1.66}^{+1.44}$ & $-0.77_{-0.90}^{+2.18}$ \\
        log\,$P_\textnormal{cloud}$(bar) & $\mathcal{U}(-6, 3)$ & $0.54_{-1.61}^{+1.52}$ & $-1.39_{-1.55}^{+0.77}$ \\
		log\,$X_\textnormal{H$_{2}$O}$ & $\mathcal{U}(-10, 0)$ & $-6.30_{-2.29}^{+2.59}$ & $-4.13_{-0.46}^{+0.63}$ \\
        log\,$X_\textnormal{V}$ & $\mathcal{U}(-10, 0)$ & $-1.61_{-0.67}^{+0.46}$ & $-3.77_{-4.21}^{+1.03}$ \\
        log\,$X_\textnormal{TiO}$ & $\mathcal{U}(-10, 0)$ & $-7.54_{-0.74}^{+0.60}$ &   $-5.95_{-0.42}^{+0.47}$ \\
        log\,$X_\textnormal{VO}$ & $\mathcal{U}(-10, 0)$ & $-8.89_{-0.68}^{+0.82}$ & $-6.72_{-1.79}^{+0.51}$ \\
		$\delta$ (ppm) & $\mathcal{U}(-10000, 10000)$ & - & $-241_{-422}^{+264}$ \\
		ln\,$\mathcal{Z}$ & - & 109.2 & 415.1 \\
		$\chi_{\nu}^{2}$ & - & 2.65 & 1.55 \\
		\hline
	\end{tabular}
    \end{threeparttable}
\end{table*}

\section{Discussion} \label{sec:Discussion}

\subsection{Stellar activity} \label{sec:stellar activity}

As mentioned in \citet{Pont2008, McCullough2014, Rackham2017, Rackham2018}, stellar activity such as unocculted spots and faculae may cause variation in transmission spectrum. As for WASP-121, several previous works discussed the impact of stellar activity to transmission spectrum. The discovery paper of \citet{Delrez2016} monitored the out-of-eclipse photometric variability from 27 non-consecutive nights between 2014 October 25 and 2014 December 8 using TRAPPIST. They found WASP-121 very quiet with standard deviation of 1.3\,mmag in their $V-$band light curves. \citet{Evans2016} accused the different radius measured at optical and NIR wavelengths to unocculted spots, and they claimed that the spot coverage should be $\sim7\%$ of the visible stellar disc. They further used the Celestron 14-inch (C14) Automated Imaging Telescope (AIT) to monitor the host star activity in two campaigns (2017 January 27 to April 23 and 2018 February 22 to April 8)~\citep[Appendx B.3]{Evans2018}, and concluded that WASP-121 had been photometrically stable over multi-week periods to $\sim5-1$\,mmag level. Further works of~\citet{Evans2018} and~\citet{Wilson2021} also suggested the stellar activity should be negligible, and it should not be responsible for the variation in transmission spectrum.

Our observation was taken during the second monitoring program of \citet{Evans2018}. As shown in their Figure~B.3, the peak-to-valley amplitudes at that time was $\sim 0.00320 \pm 0.00096$\,mag, while the period of host star activity was $6.66$\,days, $\sim80$ times larger than our relevant timescale of $\sim$ hours. Therefore, we believe that the obtained transmission spectrum of WASP-121b should be untouched by the activity of its host star.

\subsection{Composition of the atmosphere} 
\label{sec:component of WASP-121b}

Several previous works have studied the atmospheric composition of WASP-121b with LR and HR transmission spectrum. Many molecules, atoms and ions have been detected or confirmed in the atmosphere of WASP-121b (e.g H$_2$O, \ion{Ca}{I}, \ion{Ca}{II}, Cr, \ion{Fe}{I}, \ion{Fe}{II}, K, Li, Mg, Na, Ni, \ion{Sc}{II}, V, TiO and VO). However, the existence of TiO and VO is still controversial. In E16, the authors suggested the deeper transit depth in optical passband to be the evidence for the presence of TiO and VO. This was further supported in \citet{Evans2017} by the discovery of the thermal inversion layer in the dayside from their NIR thermal spectrum. However, several theory research indicated TiO/VO may not be the only absorber to induce thermal inversion. Some other metal-rich species (Fe, Mg, SiO, AlO, CaO, NaH and MgH) could also cause thermal inversion~\citep{Lothringer2018, Gandhi2020}. In addition, mechanical greenhouse effects may heat the atmosphere or suppress cooling and induce thermal inversion ~\citep{Youdin2010}. Further optical transmission spectroscopy observation of HST/STIS only obtained a detection of VO and an upper limit for TiO~(E18). High-resolution transmission spectrum of WASP-121b~\citep{Bourrier2020, Gibson2020, Cabot2020, Ben-Yami2020, Hoeijmakers2020, Borsa2021, Merritt2021, Gibson2022} have detected and confirmed Fe and V, but the existence of titanium and its oxide still remains controversial.

E18 reported the detection of VO and TiO with abundances of $-6.6_{-0.3}^{+0.2}$ and $< -7.9$, respectively. W21 obtained the mass fractions of VO and TiO as log\,$X_\textnormal{VO}$ = $-10.74_{-0.56}^{+0.49}$ and log\,$X_\textnormal{TiO}$ = $-9.98_{-0.65}^{+0.38}$, respectively. These two works are obviously not consistent. Actually, the optical transmission spectra obtained in these two works differ from each other (cf. color dots in Figure~\ref{fig:tran_spec_comp}). E18 run a series of robustness tests, including the treatment of limb-darkening coefficients and the additional parameters of GP inputs. They found that the measured transmission spectrum obtained by different instrument (G430L and G750L gratings) could not be explained by one self-consistent model. Later, W21 reported the discrepancy in optical transmission spectrum, and pointed out that the transmission spectrum may vary with time. W21 and reference therein~\citep{Parmentier2013, Komacek2020} pointed out that the temperature fluctuations might results in significant spatial and temporal variations in atmospheric constituents, and could lead to measurable variations in transmission spectrum, with predictable variation period of 50 to 100 days. The E18 observation were taken in October and November 2016, while the W21 data were obtained in January 2017, while our observation were carried out in March 2018. These datasets have a long time separation with 60 and 420 days, which is consistent with the variation timescale mentioned above. Therefore, the possibility of temporal variation of transmission spectra can not be rejected.

From our ONIR retrieval analysis, we obtain clear detection of TiO with log\,$X_\textnormal{TiO} = -5.95_{-0.42}^{+0.47}$, which is much higher than the upper limit from E18 and the abundance from W21. As for VO, its mass fraction is estimated to be $-6.72_{-1.79}^{+0.51}$\,dex in this work, consistent with that from E18 with $\sim 2.7\sigma$ confidence, and is significantly larger than that from W21. These abundance discrepancies confirm that the physical and chemical conditions in the terminator region of WASP-121b may indeed change with time. In addition, we detect a strong water absorption with the abundance of log\,$X_\textnormal{H$_{2}$O}$ = $-4.13_{-0.46}^{+0.63}$, which is between the abundance from~\citet{Evans2018}($-2.2_{-0.3}^{+0.3}$) and~\citet{Wilson2021} ($-5.05_{-0.18}^{+0.19}$). The discrepancy of H$_{2}$O abundance may come from the retrieval model assumption and the difference of optical spectrum, due to E18, W21 and this work use the same HST/WFC3 transmission spectrum from~\citet{Evans2016}.

Interestingly, as shown in the posterior distribution in Figure~\ref{fig:tran_spec_free_model}, there is actually another peak of VO in the lower abundance regime at about $-9$\,dex, consistent with that from W21. This bimodal distribution is also visible in the distribution plots and corner plots of the cloud pressure, the fraction of V and $\delta$, the offset between the HST data and the SOAR data. We find that these parameters may be coupled with each other, which might be why there are two possible solutions to deduced from the obtained transmission spectra. For the case that $\delta$ is close to zero, a large amount of V and VO are required to enhance the optical opacity to balance between the optical and the NIR transmission spectra. For the case with negative $\delta$ adopted, the HST transmission spectrum will be raised up to match better with the optical transmission spectrum, resulting in much smaller requirements for the abundances of V and VO. It is interesting to note that both E18 and W21 yielded a lower TiO and/or VO abundances, which might be due to adopting a constant $a/R_{\star}$ and $b$ (or $i$) in the optical and NIR bands for their light curve fittings. As $R_\textnormal{p}/R_{\star}$ is highly dependent on $a/R_{\star}$ and $b$, using same $a/R_{\star}$ and $b$ would result in aligning $R_\textnormal{p}/R_{\star}$ values in the optical and NIR bands. This aligning is effectively similar to applying an offset on the NIR transmission spectra to match with the optical one, just like the second case discussed above.


\section{Conclusion} \label{sec:Conclusion}

In this work, we report the study of the UHJ WASP-121b based on our ground-based 4-meter telescope SOAR/GHTS optical transit observation and literature HST NIR transmission spectra. The SOAR observations were taken on the nights of 2018 February 7 and March 1, but the first night only covered partial transit due to weather loss. As a result, only data taken in the second night was employed in this study. By dividing the stellar spectra in 19 passbands, we extracted 19 spectroscopic transit light curves and fit them using Gaussian process method. Then we derive transit depths in 19 spectral bins, i.e the transmission spectrum of WASP-121b. The stellar activity of the host star is weak with variation of $0.00320 \pm 0.00096$\,mag over a period of $\sim6.6$\,days from the simultaneous photometric monitoring program by ~\citet{Evans2018}, and thus should not produce unignorable imprint on our obtained transmission spectrum with a timescale of $\sim$hours. 

The derived optical transmission spectrum shows an increased slope toward blue wavelength with a slope $\alpha = -9.95_{-1.49}^{+1.47}$, much smaller than the hydrogen-dominated Rayleigh scattering ($\alpha = -4$), which could not be explained by Rayleigh scattering alone. There are notable differences between our spectrum and those in the literature from \citet{Evans2018, Wilson2021}, which implies that the atmosphere of WASP-121b may experience temporal variation. Note that our data analysis procedures passed our validation experiment, by the small offset $\sim472 \pm 286 \pm 25 $\,ppm between the GMOS transmission spectra derived by us from the 1D spectra provided by Dr. Jamie Wilson and Dr. Neale Gibson via private communication with those from \citet{Wilson2021}. This small difference is caused by the different setting in orbital parameters during the light curve analysis.

Retrieval analysis are performed on the optical data and the combination of the optical data and literature HST/WFC3 NIR data, to evaluate the composition and temperature structure of WASP-121b. Both analysis support the presence of thermal inversion layer in the day-side atmosphere, with turn-over point parameters consistent with each other within their quoted uncertainties. The best model is the one obtained on the combined data set considering TiO, VO, H$_2$O and atom V, which yields a clear detection of TiO with log\,$X_\textnormal{TiO} = -5.95_{-0.42}^{+0.47}$, much higher than those from E18 and W21. VO is also detected with the median log\,$X_\textnormal{VO} = -6.72$ plus another peak at $\sim -9$. While the median value is consistent with that from E18, the secondary peak is in line with that from W21. Although other species like SiO and mechanical greenhouse effects can also induce thermal inversion, the transmission data collected in this work favours the presence of TiO in WASP-121b, as discussed in Section~\ref{sec:component of WASP-121b}.

We argue that some parameters particularly the NIR-to-OPT offset $\delta$, the cloud deck pressure, and abundances of VO and V may be coupled, as illustrated by the double clumps seen in the corner distribution of the above-mentioned parameters in Fig.~\ref{fig:free_model_HST_corner}. If fixing $\delta$ to zero, more V and VO are required to enhance the opacity in the optical to account for the discrepancy between the optical and the NIR transmission spectra. On the other side, setting a negative $\delta$, which raises up the HST transmission spectra, results in much smaller requirements for the contribution from V and VO. Therefore, we caution that putting $\delta$ into retrieval may induce further uncertainties.  

To summarize, the discrepancies seen in the optical transmission spectra and the determined abundances of VO, TiO and H$_2$O between this work and those from literature provide direct evidence of the variation of WASP-121b's atmosphere, as suggested by ~\citet{Wilson2021}. According to their work and the reference therein, the $T-P$ profile of WASP-121b atmosphere may lie near the condensation curves of a number of species, which may result in significant temporal variations in the atmospheric components. Therefore, we strongly recommend that multi-epoch low and high-resolution transit observations should be conducted for WASP-121b to study the temporal variation in its transmission spectrum.

\normalem
\begin{acknowledgements}
We thank the anonymous reviewer for their constructive comments. This research is supported by the National Key R\&D Program of China No. 2019YFA0405102 and 2019YFA0405502, the National Natural Science Foundation of China under grants No. 42075123, 62127901, 11988101, 42005098, 12073044. This work is also supported by the China Manned Space Project with NO. CMS-CSST-2021-B12. MZ, YQS are supported by the Chinese Academy of Sciences (CAS), through a grant to the CAS South America Center for Astronomy (CASSACA) in Santiago, Chile.

This work is based on observations obtained at the Southern Astrophysical Research (SOAR) telescope, which is a joint project of the Minist\'{e}rio da Ci\^{e}ncia, Tecnologia e Inova\c{c}\~{o}es (MCTI/LNA) do Brasil, the US National Science Foundation’s NOIRLab, the University of North Carolina at Chapel Hill (UNC), and Michigan State University (MSU).

\end{acknowledgements}

\section{Appendix A: additional figures}
\renewcommand{\thetable}{A\arabic{table}}
\renewcommand{\thefigure}{A\arabic{figure}}
\setcounter{figure}{0}

\begin{figure*}
	\includegraphics[width=\textwidth]{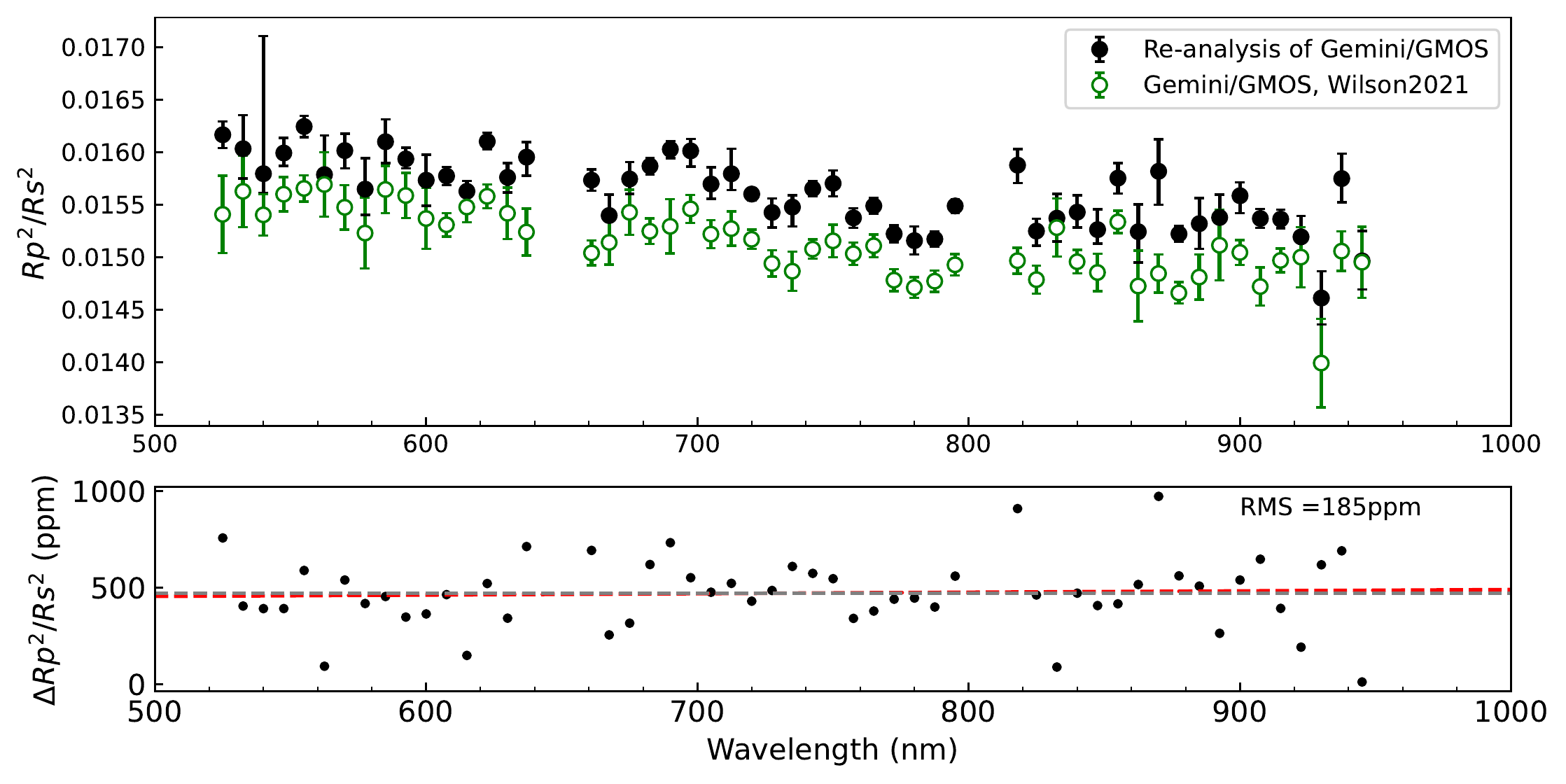}
    \caption{\textit{Top}: The optical transmission spectrum of WASP-121b obtained by Gemini/GMOS, The black circle dots with errorbars are the Gemini/GMOS transmission spectrum re-analysed by us, while the green circle dots are the transmission spectrum published in~\citet{Wilson2021}. Both results are well consistent. \textit{Bottom}: The $\Delta R_{p}^{2}/R_{s}^{2}$ of the two transmission spectrum. The dashed grey line is the mean value of $\Delta R_{p}^{2}/R_{s}^{2} = 472$\,ppm, and the dashed red line is a simple linear polynomial function to fit these $\Delta R_{p}^{2}/R_{s}^{2}$, which is very close to the mean value.}
    \label{fig:tran_spec_gmos}
\end{figure*}

\begin{figure*}
	\includegraphics[width=\columnwidth]{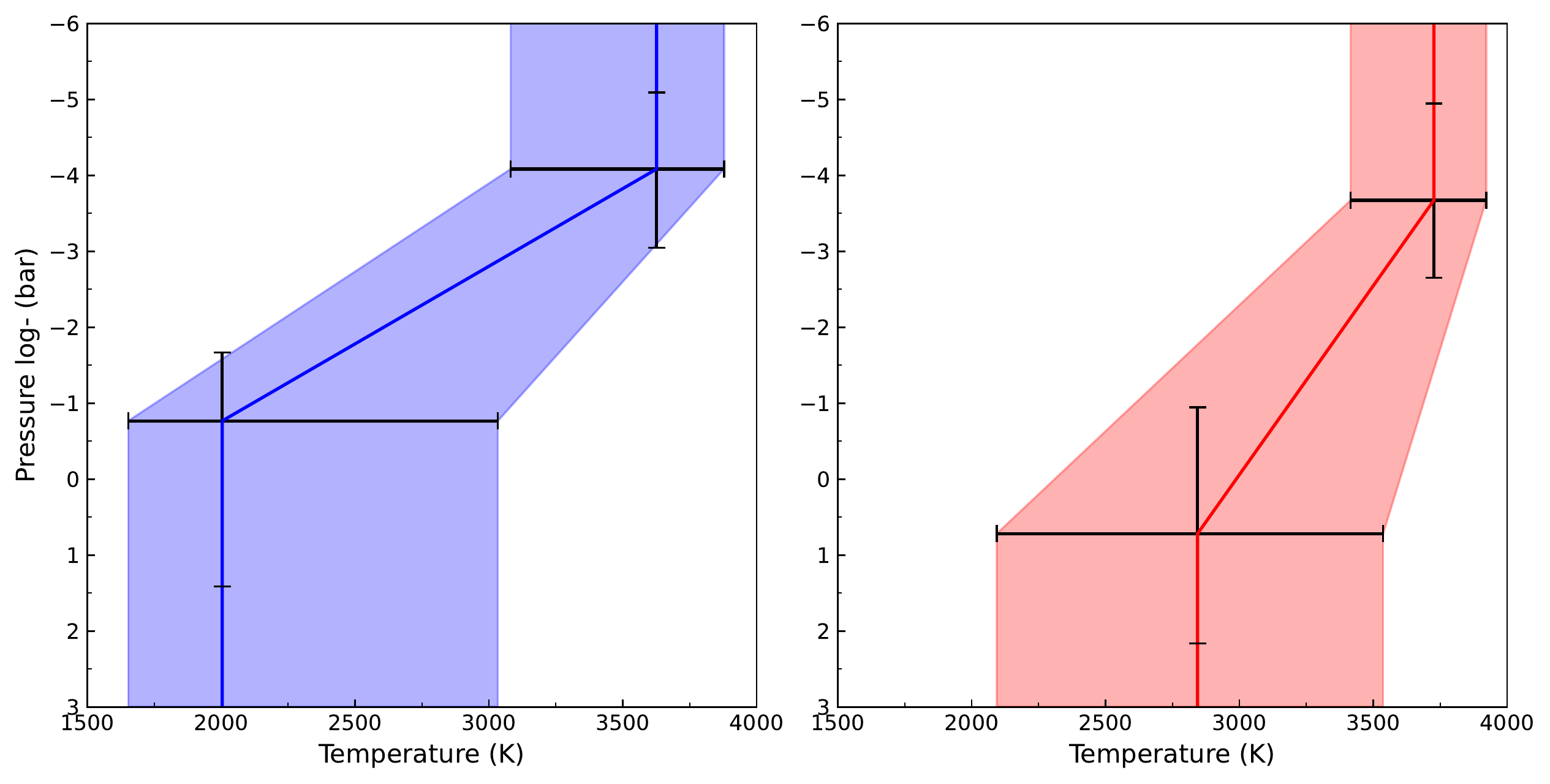}
    \caption{\textit{Left}: The $T-P$ profile retrieved from OPT data and the $1\sigma$ confidence interval. \textit{Right}: The $T-P$ profile retrieved from ONIR data and the $1\sigma$ confidence interval.}
    \label{fig:TP_profile}
\end{figure*}

\begin{figure*}
	\includegraphics[width=\textwidth]{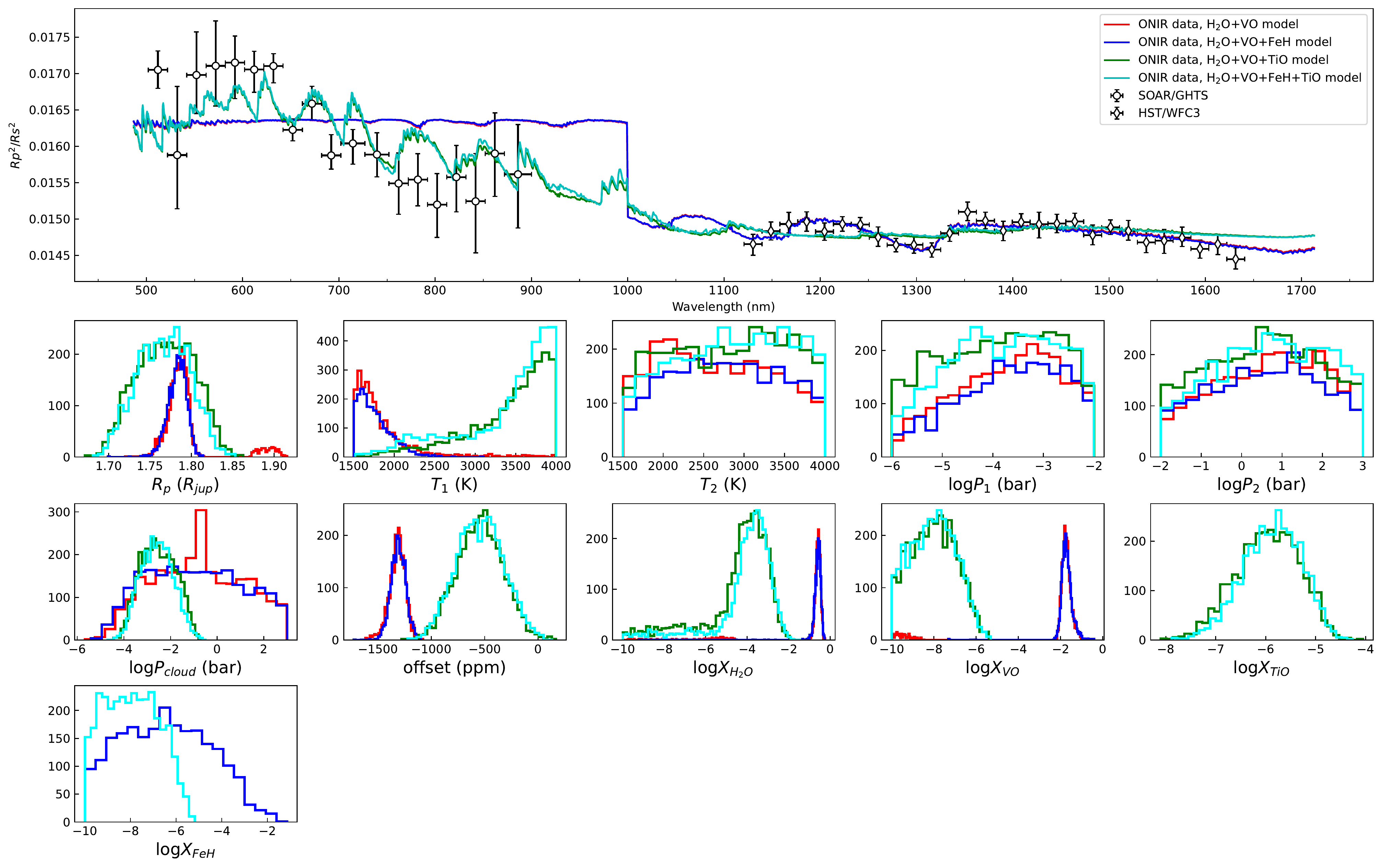}
    \caption{Same as Fig.~\ref{fig:tran_spec_free_model}, the retrieval models and posterior distributions of the first step in retrieval.}
    \label{fig:tran_spec_model_A}
\end{figure*}

\begin{figure*}
	\includegraphics[width=\textwidth]{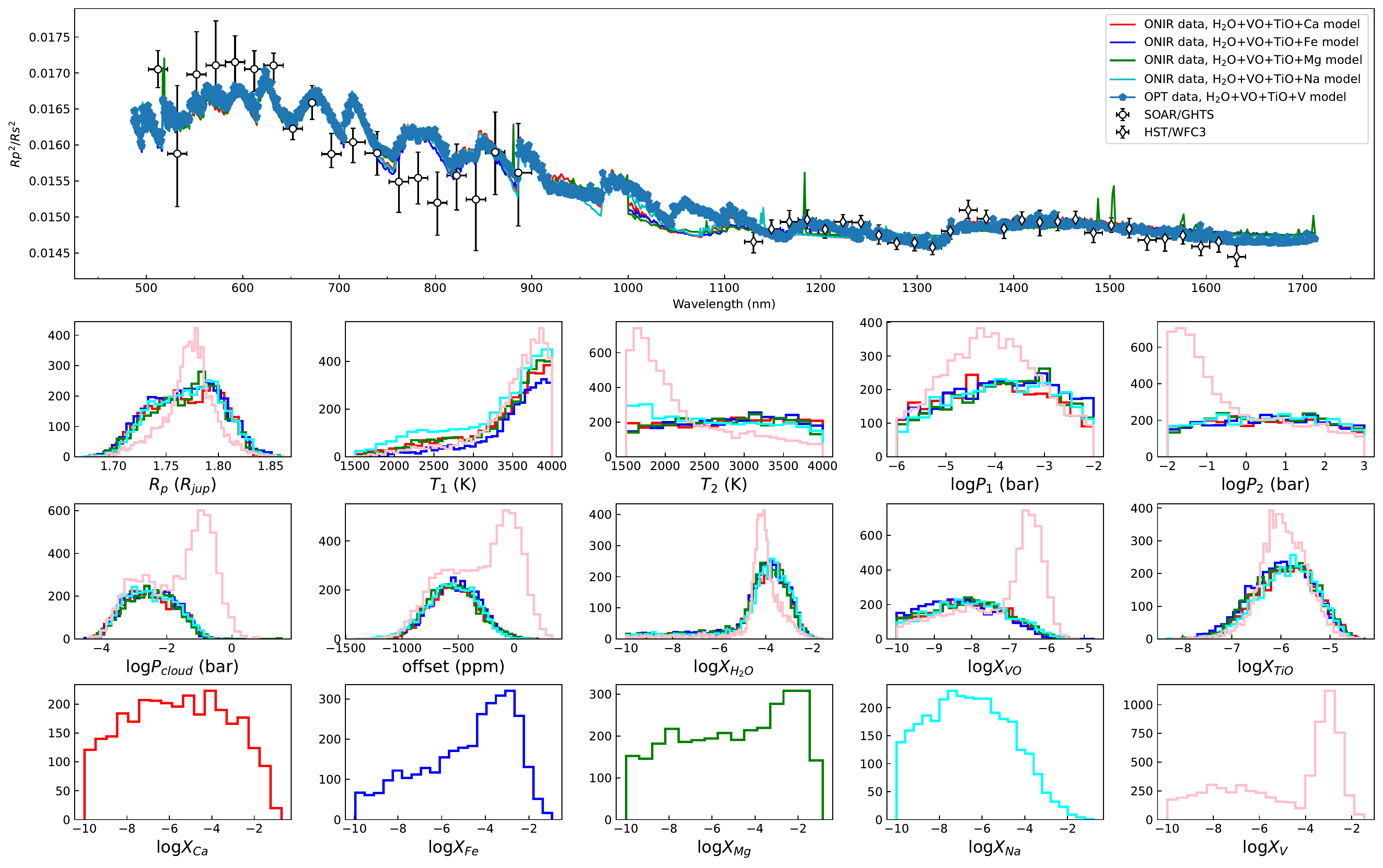}
    \caption{Same as Fig.~\ref{fig:tran_spec_free_model}, the retrieval models and posterior distributions of the second step in retrieval.}
    \label{fig:tran_spec_model_B}
\end{figure*}

\begin{figure*}
	\includegraphics[width=\textwidth]{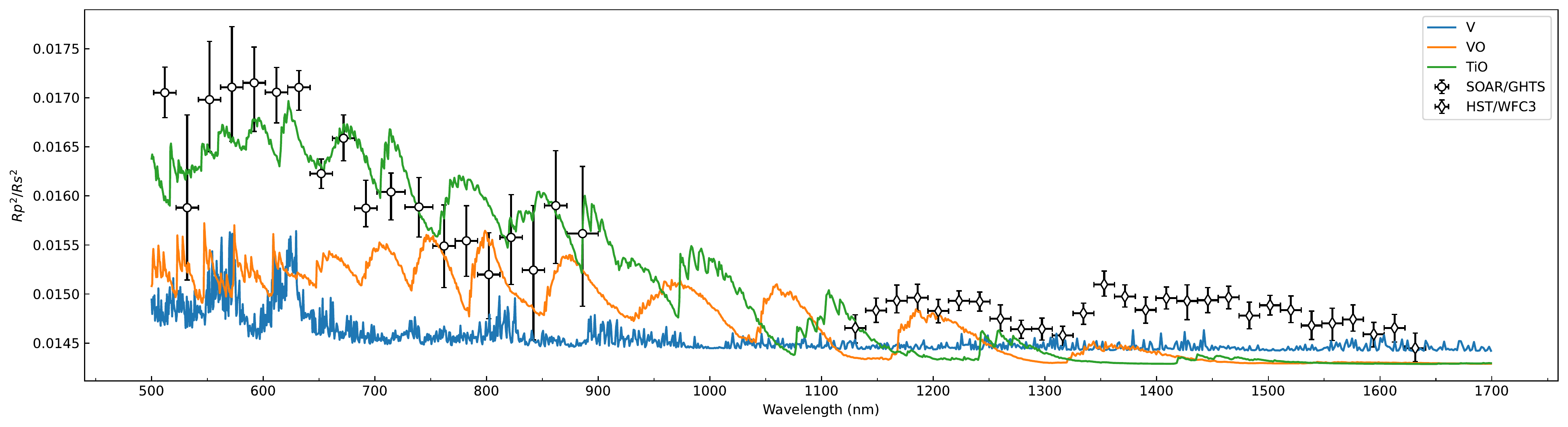}
    \caption{The OPT and ONIR transmission spectrum of WASP-121b with the contribution of V, VO and TiO.}
    \label{fig:tran_spec_species_depth}
\end{figure*}

\begin{figure*}
	\includegraphics[width=\textwidth]{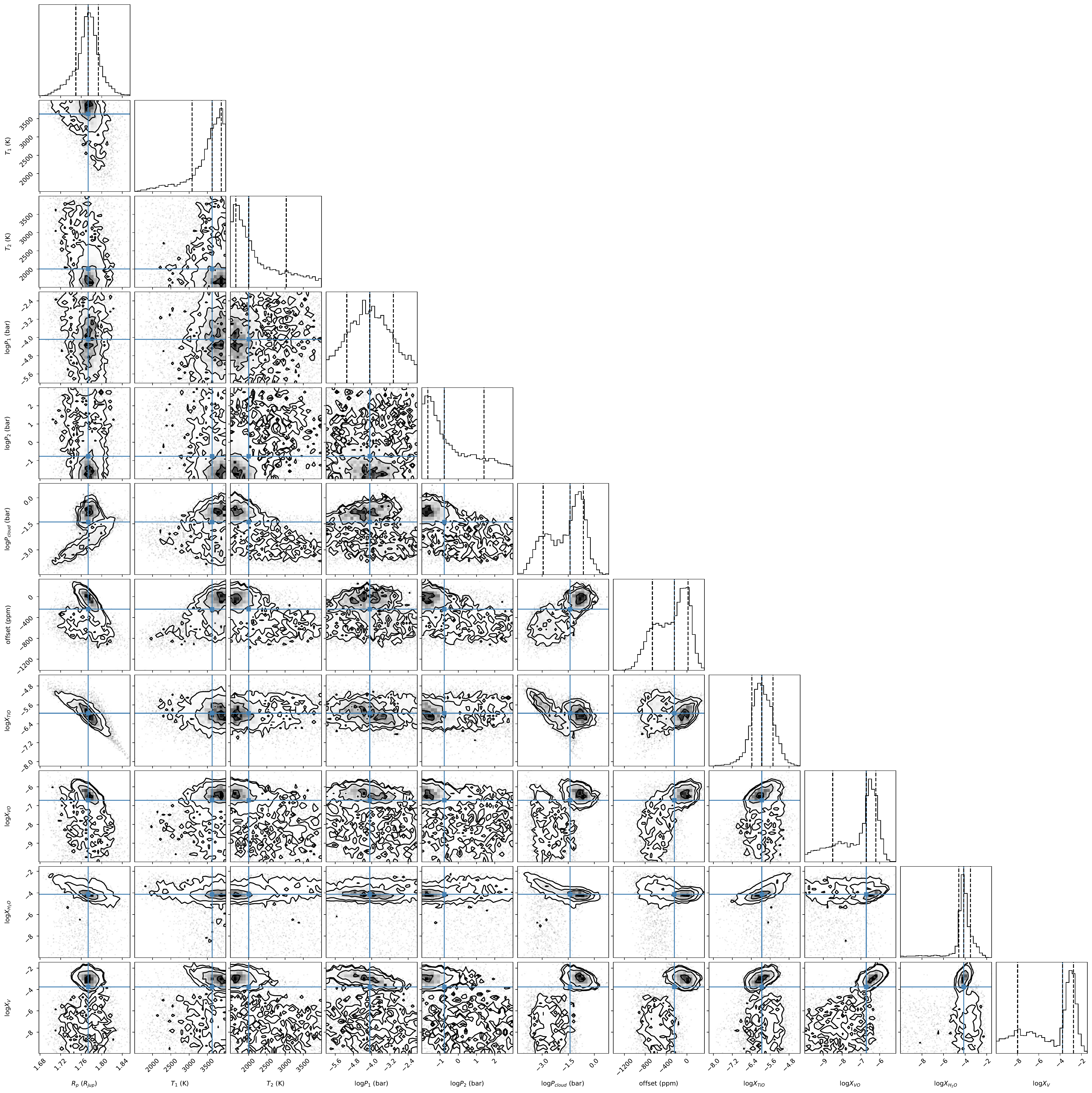}
    \caption{The posterior distribution corner plot of The B5 model in Table~\ref{tab:retrieval_statistics}.}
    \label{fig:free_model_HST_corner}
\end{figure*}
  
\bibliographystyle{raa}
\bibliography{paper_final}

\end{document}